\documentclass[aps,bibnotes,showpacs,superscriptaddress]{revtex4}

\usepackage{graphicx}
\usepackage{amssymb}
\usepackage{amsmath}
\usepackage{amsfonts}
\usepackage[dvipdfm]{hyperref}
\usepackage{bm}

\DeclareMathAccent{\ring}{\mathalpha}{operators}{"17}
\providecommand{\st}[1]{_{\text{#1}}}

\def\bra{\ensuremath{\langle}}
\def\ket{\ensuremath{\rangle}}

\def\kv{\bv{k}}

\def\rv{\bv{r}}

\def\b0{\bv{0}}

\newcommand{\bitem}{\begin{itemize}}
\newcommand{\eitem}{\end{itemize}}
\newcommand{\benum}{\begin{enumerate}}
\newcommand{\eenum}{\end{enumerate}}
\newcommand{\bblock}[1]{\begin{block}{#1}}
\newcommand{\eblock}{\end{block}}
\newcommand{\bmini}[1]{\begin{minipage}{#1}}
\newcommand{\emini}{\end{minipage}}
\newcommand{\btab}[1]{\begin{tabular}{#1}}
\newcommand{\etab}{\end{tabular}}
\newcommand{\btabn}[1]{\begin{tabular}{#1}}
\newcommand{\etabn}{\end{tabular}}
\newcommand{\beq}{\begin{equation}}
\newcommand{\eeq}{\end{equation}}
\newcommand{\beqn}{\begin{equation*}}
\newcommand{\eeqn}{\end{equation*}}
\newcommand{\bmult}{\begin{multline}}
\newcommand{\emult}{\end{multline}}
\newcommand{\bsplit}{\begin{split}}
\newcommand{\esplit}{\end{split}}

\newcommand{\bv}[1]{\mathbf{#1}}

\begin{document}

\title{Spreading Dynamics of Nanodrops: A Lattice Boltzmann Study}

\author{Markus Gross}
\affiliation{Interdisciplinary Centre for Advanced Materials Simulation (ICAMS), Ruhr-Universit\"at Bochum\\ Universit\"atsstra{\ss}e 150\\ 44780 Bochum, Germany}

\author{Fathollah Varnik}
\affiliation{Interdisciplinary Centre for Advanced Materials Simulation (ICAMS), Ruhr-Universit\"at Bochum\\ Universit\"atsstra{\ss}e 150\\ 44780 Bochum, Germany}

\begin{abstract}
Spreading of nano-droplets is an interesting and technologically relevant phenomenon where thermal fluctuations lead to unexpected deviations from well-known deterministic laws. Here, we apply the newly developed fluctuating non-ideal lattice Boltzmann method [Gross et al., J Stat Mech, P03030 (2011)] for the study of this issue. Confirming the predictions of Davidovich and coworkers [PRL {\bf 95}, 244905 (2005)], we provide the first independent evidence for the existence of an asymptotic, self-similar noise-driven spreading regime in both two- and three-dimensional geometry. The cross over from the deterministic Tanner's law, where the drop's base radius $b$ grows (in 3D) with time as $b \sim t^{1/10}$ and the noise dominated regime where $b \sim t^{1/6}$ is also observed by tuning the strength of thermal noise.

\end{abstract}



\maketitle

\section{Spreading of nano-droplets}
\begin{figure}[b]\centering
   \includegraphics[width=0.3\linewidth]{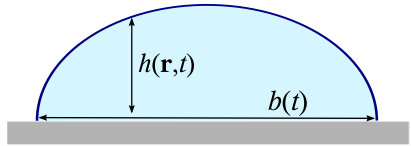}
   \caption{Geometry of a spreading droplet on a solid substrate. $h(\rv,t)$ is the local height, $b(t)$ the base diameter and $\rv$ is the coordinate in the plane of the substrate.}
    \label{fig:spread_geom}
\end{figure}

The spreading of a viscous liquid droplet on a solid substrate is a fundamental and long-studied phenomenon \cite{deGennesQuere_book,deGennes_review1985,bonn_review2009} of relevance to many applications, such as, for example, painting or coating techniques. If a small macroscopic droplet is placed on a perfectly wetting substrate, one expects that, after an initial transition period, the base radius (see Fig.~\ref{fig:spread_geom}) increases according to a power-law, $b(t)\propto t^n$, with an exponent $n=1/10$ in 3D and $n=1/7$ in 2D. This relation can be derived from lubrication theory \cite{safran_book,deGennesQuere_book} and is known as \textit{Tanner's law} \cite{tanner1979}.

The experimentally observed value of the exponent $n$ often deviates from the theoretical predictions and is found to depend on several physical parameters, such as substrate roughness, liquid viscosity or vapor density. This is reflected in the range of values that have been reported by experiments \cite{marmur1983} and simulations \cite{raiskinmaeki_2000,dupuis_spreading_2003,iwahara_2003}.

Most experiments deal with the spreading of droplets that are at least of micrometer size and thus are unaffected by thermal fluctuations. It is, however, interesting to ask what happens to droplets at the nanoscale, where thermal fluctuations become important. Recently, the spreading of droplets under the influence of thermal fluctuations has been investigated numerically and theoretically \cite{davidovitch_spreading_2005}.
There, a stochastic lubrication equation of the form
\beq \partial_t h = -\frac{\sigma}{3\eta} \nabla\cdot (h^3 \bv{\nabla} \nabla^2 h) + \sqrt{\frac{2 k_B T}{3\eta}} \nabla\cdot \left[ h^{3/2} \xi \right]
\label{lubri}
\eeq
has been proposed for the evolution of the film height $h$ of the spreading droplet (see also \cite{gruen_2006}). Here, $\sigma$ is the surface tension, $\eta$ the dynamic viscosity, and $\xi$ is a Gaussian white noise with unit variance,
\beq \bra \xi(\rv,t)\xi(\rv',t')\ket = \delta(\rv-\rv')\delta(t-t')\,.\label{lubri_noise}
\eeq
Since the total volume $V$ of the droplet is conserved, one additionally requires
\beq V=\int d\rv h(\rv,t)=\text{const.}\,\eeq
The lubrication equation \eqref{lubri}, which is a fully nonlinear stochastic differential equation with multiplicative noise, can be derived from the Navier-Stokes equations for a liquid film amended with the usual Landau-Lifshitz random stress tensor \cite{gruen_2006}. The prefactor of the noise in eq.~\eqref{lubri} ensures that, for a flat film in equilibrium, $h$ obeys the standard capillary wave fluctuation spectrum [$\bra |h(\kv)|^2\ket = k_B T / \sigma k^2$, with $k$ being the surface wavevector].
The first term on the r.h.s.\ of eq.~\eqref{lubri} arises from the driving force due to Laplace pressure, $\sigma \nabla^2 h$, and is the only term governing the deterministic evolution of a macroscopic droplet in the lubrication approach.

Comparing the factor $h^3$ in front of the surface tension term with $h^{3/2}$ in front of the noise shows that the relative importance of thermal noise increases with decreasing film height (cf.\ \cite{eggers_jet2002,gruen_2006,fetzer_dewetting_2007}).
Thus, one expects that, at late times, where the droplet is sufficiently flat, the evolution will always be dominated by thermal noise.
Remarkably, in this case, a significant increase in the long-time spreading rate, associated with a change of the exponent in Tanner's law has been derived analytically and by direct numerical solution of eq.~\eqref{lubri} \cite{davidovitch_spreading_2005}.

However, these predictions have so far not been confirmed experimentally or by simulation of the full fluid dynamical equations.
Presently available Molecular Dynamics simulations of spreading nano-droplets could not reach far enough into the asymptotic regime to arrive at definite conclusions, although some agreements with the predictions of the stochastic lubrication equation were observed \cite{willis_freund_jpcm2009}. Interestingly, an earlier Monte-Carlo study on polymer droplets found Tanner's classic law to hold \cite{milchev_binder_jcp2002}.
Here, results of Lattice Boltzmann (LB) simulations will be presented confirming the existence of a noise-enhanced spreading regime as predicted by the stochastic lubrication equation \eqref{lubri}.

Regarding the origin of deviations from Tanner's law, we note that in the limit where either the surface tension or the stochastic driving force dominates, eq.~\eqref{lubri} admits for a non-trivial scaling solution of the form \cite{davidovitch_spreading_2005}
\beq h(\rv,t) = |\rv|^{-\beta} f(|\rv|/t^z)\,. \label{lubri_scale_sol}\eeq
In conjunction with volume conservation, the scaling ansatz implies that $\beta=d$ to ensure time-independence of the volume.
Since $h$ is a random variable, a suitable measure of the lateral droplet size can be defined as the variance of the height profile,
\beq s(t) = \left[\frac{1}{V} \left\bra \int d\rv |\rv-\bv{x}_0|^2 h(x,t) \right\ket\right]^{1/2} \propto t^{\frac{3+\beta}{2z}}\propto t^n \,,\label{lubri_size}\eeq
where $\bv{x}_0=\int d\rv \,\rv h(\rv,t) /V$ is the instantaneous position of the center of the droplet and the proportionalities follow by using the scaling ansatz, eq.~\eqref{lubri_scale_sol}.
Inserting the scaling ansatz into eq.~\eqref{lubri} and requiring form-invariance of the resulting equation shows that, in case where the surface tension term dominates, the droplet size $s$ follows Tanner's classical law, $s=t^n$ with $n=1/7$ (2D) and $n=1/10$ (3D), whereas, for the noise-dominated regime one obtains $n=1/4$ (2D) and $n=1/6$ (3D), thus, enhanced spreading.
The conditions for the dominance of the noise in the spreading dynamics are estimated in \cite{davidovitch_spreading_2005} as 
\begin{equation}
\begin{split} 
h/h_0 &\ll l_T^{1/3}/W^{1/6}h^{1/6}\qquad \text{(2D),}\\
h/h_0 &\ll l_T^{1/3}/h_0^{1/3}\qquad \text{(3D),}
\end{split}
\label{noise_cond}
\end{equation}
where $h_0$ is the initial droplet height, $W$ is the width of the drop in the 2D case ($W=1$ lattice units in the present setup) and $l_T=\sqrt{k_B T / \sigma}$ is the thermal length.

\begin{figure}[t]\centering
   (a)\includegraphics[width=0.45\linewidth]{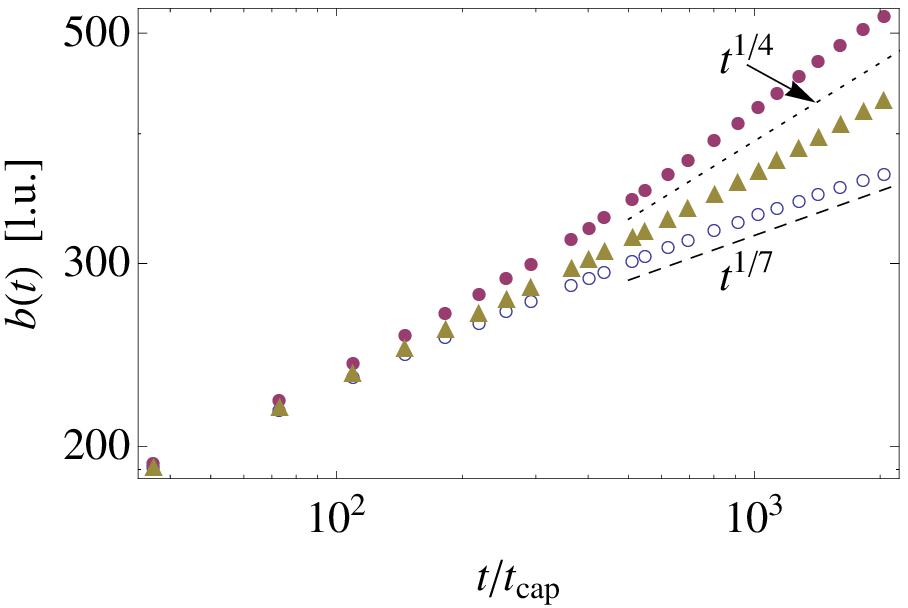}
   (b)\includegraphics[width=0.45\linewidth]{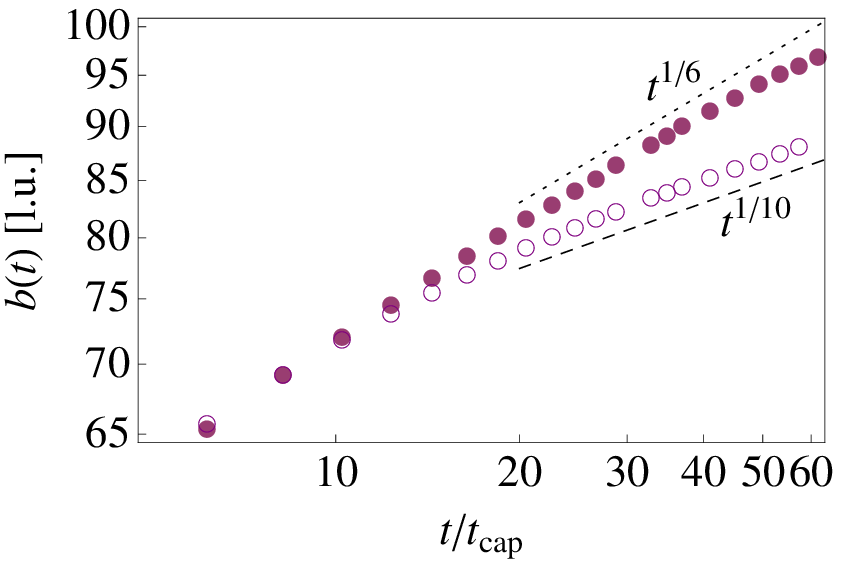}
   (c)\includegraphics[width=0.45\linewidth]{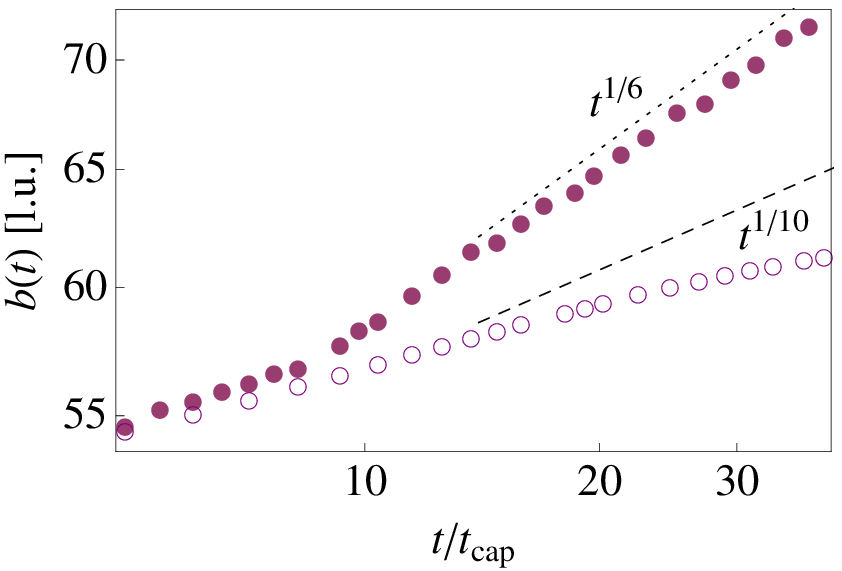}
   \caption{Spreading of a droplet on a solid substrate in 2D (a) and 3D (b,c). In each panel, the time evolution of the base radius for different noise intensities is shown. The dashed line represents the deterministic Tanner law, while the dotted line represents the noise-dominated law. Symbols: $\circ$ no thermal noise, $\blacktriangle$ $k_B T=4\times 10^{-8}$, $\bullet$ $k_B T=10^{-7}$ in (a), $5\times 10^{-6}$ in (b) and $10^{-5}$ in (c). Simulation parameters (all in l.u.): densities $\rho_L=1.0$, $\rho_V=0.1$, surface tension $\sigma=0.0027$, interface width $\approx 5$, viscosity $\eta/\rho_{L,V}=0.167$ in (a), $0.067$ in (b) and $0.167$ in (c) in 3D. $t\st{cap}=\eta R_0/\sigma$ is the capillary time and $R_0$ the initial base radius of the droplet.}
    \label{fig:spread_rad}
\end{figure}

\begin{figure}[t]\centering
   (a)\includegraphics[width=0.45\linewidth]{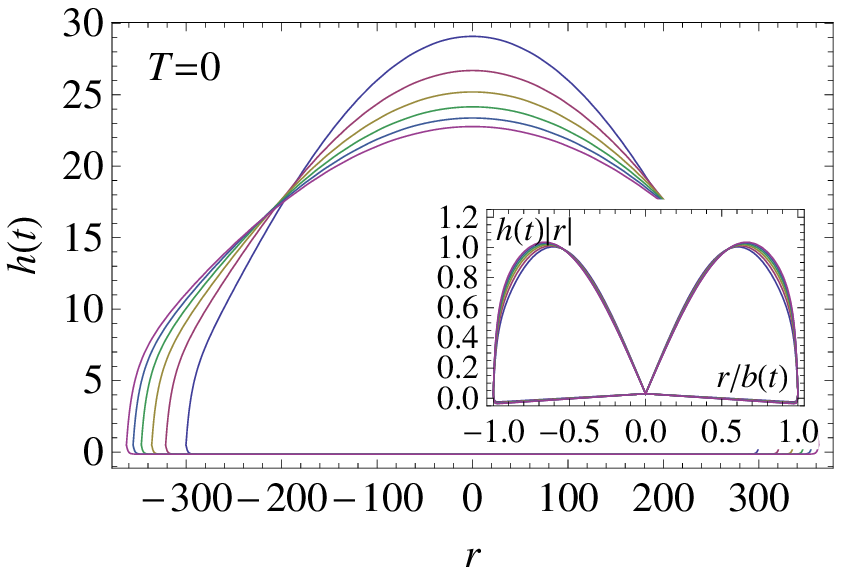}
   (b)\includegraphics[width=0.45\linewidth]{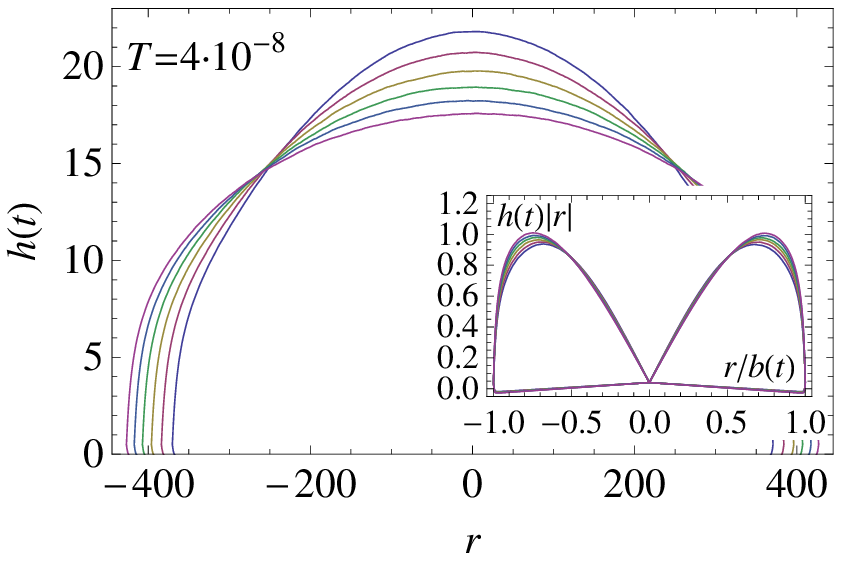}
   \caption{Instantaneous droplet profiles in the self-similar spreading regime in 2D, (a) without and (b) with noise. The insets show the rescaled droplet profiles (the rescaled height $h(t)|r|$ is normalized to one). Simulation parameters are as in Fig.~\ref{fig:spread_rad}a.}
    \label{fig:spread_prof}
\end{figure}

\section{Simulation method and results}

In order to study nanofluidic phenomena, we have recently introduced thermal fluctuations into a number of existing non-ideal fluid models auch as the Lee-Fischer model \cite{gross_fdbe_2011}, the Shan-Chen model \cite{gross_dsfd2010} and the modified-equilibrium model \cite{gross_flbe_2010}. For the details on the deterministic versions of the above mentioned models, we refer to\cite{lee_fischer_2006,shan_Lattice_1993,shan_chen_1994,swift_prl1995,swift_pre1996}.

Using the Lee-Fisher version of our implementation, we study the spreading dynamics of nanodrops both in two and three dimension. Similar results have also been obtained with the fluctuating modified equilibrium method.
For each simulation, the initial configuration is a half-spherical droplet of radius $R_0=90$~(all quantities are given in lattice units [l.u.]). The size of the simulation domain is $L_X\times L_Y=1200\times 110$, with periodic boundary conditions along the $x$- and solid wall boundary conditions along the $y$-axis. The interface width is approximately 5, the surface tension is $0.0027$, the liquid and vapor densities are $\rho_L=1.0$, $\rho_V=0.1$ and the relaxation times are set to $\tau=1.0$.
It is found that, in the present case, both the base radius $b(t)$ and the variance of the height profile, $s(t)$ [eq.~\eqref{lubri_size}], give a completely equivalent measure of the droplet size. Furthermore, since the noise intensities are weak compared to the surface energy, no averaging is necessary to improve the statical accuracy and the plots are thus each obtained from a single simulation.
Owing to the slowness of the spreading process, the data had to be acquired over more than $10^7$ timesteps.
For the present setup, the conditions of eq.~\eqref{noise_cond} predict that noise dominates when $h/h_0 \lesssim 0.1$, in good agreement with our simulations.

In Fig.~\ref{fig:spread_rad}, the time evolution of the base radius $b$ is shown in two (a) and three (b,c) dimensions, both with and without thermal noise. It is observed that for the noiseless spreading (open symbols), the base radius asymptotically follows the classical Tanner law. Similar to the experimental case, the exponent in the deterministic case is found to be quite sensitive to the choice of simulation parameters, as can be seen in Fig.~\ref{fig:spread_rad}c, where a viscosity by a factor of 2.5 larger (both for liquid and vapor) than in case (b) is used. 
On the other hand, if thermal noise is present, the base radius asymptotically follows the spreading law expected from the stochastic lubrication equation \eqref{lubri}. Interestingly, as Fig.~\ref{fig:spread_rad}c shows, this behavior is quite robust and much less dependent on the simulation parameters than in the deterministic case
(e.g., in Fig.~\ref{fig:spread_rad}a, only a slight deviation between the curves for different values of $k_B T$ is visible).

Fig.~\ref{fig:spread_prof} shows the interface profile of a spreading droplet in the late-time regime in 2D. The self-similar character of the droplet evolution can be explicitly seen by rescaling the profile according to the ansatz of eq.~\eqref{lubri_scale_sol} (in the original work \cite{davidovitch_spreading_2005}, $h$ was multiplied by $b(t)$ instead of $|r|$, which is equally well suited). As shown in the insets of Fig.~\ref{fig:spread_prof}, the rescaled profiles collapse onto a master curve, which essentially represents the scaling function $f$ of eq.~\eqref{lubri_scale_sol}.

It is noteworthy that, generally, we find 3D simulations to agree less well with Tanners law if the dynamic viscosity of the vapor phase is enhanced. This is largely independent of the viscosity ratio between liquid and vapor. In the literature, it seems to be not completely clear what precisely the effect of the vapor viscosity is on spreading. Additional studies and extensive simulations are necessary in order to arrive at definite conclusions on this issue.

\section{Summary}
In this work the newly developed fluctuating non-ideal fluid LB model has been applied to the study of the spreading dynamics of nanodrops on perfectly flat solid surfaces. So far, Molecular Dynamics studies could not reach the asymptotic, self-similar spreading regime. Our simulations provide the first independent evidence for the existence of an asymptotic self-similar spreading regime dominated by thermal noise \cite{davidovitch_spreading_2005}. By tuning the magnitude of the noise term, we could observe the cross over from the classical Tanner's law, where the drop's base radius $b$ grows with time as $b \sim t^{1/10}$ and the noise dominated regime where $b \sim t^{1/6}$. The observed scaling exponent is found to be more robust (regarding parameter selection) in the noise dominated case than for purely deterministic dynamics.

Presently, only the effect of thermal fluctuations on spreading has been studied. However, it is well known that fluids at the nanoscale are also strongly influenced by van der Waals interactions (disjoining pressure). It was argued in \cite{davidovitch_spreading_2005} that, for complex fluids, these effects are negligible at least in a certain parameter region due to their weaker van der Waals interactions. Furthermore, complex fluids are usually governed by non-Newtonian constitutive relations and it is not clear how these effects will influence spreading. This will be an interesting aspect to study in the future.

\section*{Acknowledgments}
M.G. gratefully acknowledges the grant provided by the Deutsche Forschungsgemeinschaft (DFG) under the number Va 205/5-2. ICAMS gratefully acknowledges funding from ThyssenKrupp AG, Bayer MaterialScience AG, Salzgitter Mannesmann Forschung GmbH, Robert Bosch GmbH, Benteler Stahl/Rohr GmbH, Bayer Technology Services GmbH and the state of North-Rhine Westphalia as well as the European Commission in the framework of the European Regional Development Fund (ERDF).


\begin{thebibliography}{10}
\expandafter\ifx\csname url\endcsname\relax
  \def\url#1{\texttt{#1}}\fi
\expandafter\ifx\csname urlprefix\endcsname\relax\def\urlprefix{URL }\fi
\providecommand{\bibinfo}[2]{#2}
\providecommand{\eprint}[2][]{\url{#2}}

\bibitem{deGennesQuere_book}
\bibinfo{author}{de~Gennes, P.~G.}, \bibinfo{author}{Brochard-Wyart, F.} \&
  \bibinfo{author}{Quere, D.}
\newblock \emph{\bibinfo{title}{{Capillarity and Wetting Phenomena: Drops,
  Bubbles, Pearls, Waves}}} (\bibinfo{publisher}{Springer},
  \bibinfo{year}{2003}).

\bibitem{deGennes_review1985}
\bibinfo{author}{de~Gennes, P.~G.}
\newblock \bibinfo{title}{Wetting: statics and dynamics}.
\newblock \emph{\bibinfo{journal}{Rev. Mod. Phys.}}
  \textbf{\bibinfo{volume}{57}}, \bibinfo{pages}{827} (\bibinfo{year}{1985}).

\bibitem{bonn_review2009}
\bibinfo{author}{Bonn, D.}, \bibinfo{author}{Eggers, J.},
  \bibinfo{author}{Indekeu, J.}, \bibinfo{author}{Meunier, J.} \&
  \bibinfo{author}{Rolley, E.}
\newblock \bibinfo{title}{Wetting and spreading}.
\newblock \emph{\bibinfo{journal}{Rev. Mod. Phys.}}
  \textbf{\bibinfo{volume}{81}}, \bibinfo{pages}{739} (\bibinfo{year}{2009}).

\bibitem{safran_book}
\bibinfo{author}{Safran, S.~A.}
\newblock \emph{\bibinfo{title}{Statistical Thermodynamics of Surfaces,
  Interfaces and Membranes}} (\bibinfo{publisher}{Addison-Wesley Publishing},
  \bibinfo{year}{1994}), \bibinfo{edition}{1st} edn.

\bibitem{tanner1979}
\bibinfo{author}{Tanner, S.~H.}
\newblock \bibinfo{title}{The spreading of silicone oil drops on horizontal
  surfaces}.
\newblock \emph{\bibinfo{journal}{J. Phys. D}} \textbf{\bibinfo{volume}{12}},
  \bibinfo{pages}{1473} (\bibinfo{year}{1979}).

\bibitem{marmur1983}
\bibinfo{author}{Marmur, A.}
\newblock \bibinfo{title}{Equilibrium and spreading of liquids on solid
  surfaces}.
\newblock \emph{\bibinfo{journal}{Adv. Colloid Interface Sci.}}
  \textbf{\bibinfo{volume}{19}}, \bibinfo{pages}{75} (\bibinfo{year}{1983}).

\bibitem{raiskinmaeki_2000}
\bibinfo{author}{Raiskinmaeki, P.}, \bibinfo{author}{Koponen, A.},
  \bibinfo{author}{Merikoski, J.} \& \bibinfo{author}{Timonen, J.}
\newblock \bibinfo{title}{Spreading dynamics of three-dimensional droplets by
  the lattice-{B}oltzmann method}.
\newblock \emph{\bibinfo{journal}{Comp. Mat. Sci.}}
  \textbf{\bibinfo{volume}{18}}, \bibinfo{pages}{7} (\bibinfo{year}{2000}).

\bibitem{dupuis_spreading_2003}
\bibinfo{author}{Dupuis, A.}, \bibinfo{author}{Briant, A.~J.},
  \bibinfo{author}{Pooley, C.~M.} \& \bibinfo{author}{Yeomans, J.~M.}
\newblock \bibinfo{title}{Droplet spreading on heterogeneous surfaces using a
  three-dimensional lattice {B}oltzmann model}.
\newblock In \bibinfo{editor}{Sloot, P. M.~A.} \emph{et~al.} (eds.)
  \emph{\bibinfo{booktitle}{ICCS 2003, LNCS 2657}}, \bibinfo{pages}{1024}
  (\bibinfo{publisher}{Springer-Verlag Berlin Heidelberg},
  \bibinfo{year}{2003}).

\bibitem{iwahara_2003}
\bibinfo{author}{Iwahara, D.}, \bibinfo{author}{Shinto, H.},
  \bibinfo{author}{Miyahara, M.} \& \bibinfo{author}{Higashitani, K.}
\newblock \bibinfo{title}{Liquid drops on homogeneous and chemically
  heterogeneous surfaces: A two-dimensional lattice {B}oltzmann study}.
\newblock \emph{\bibinfo{journal}{Langmuir}} \textbf{\bibinfo{volume}{19}},
  \bibinfo{pages}{9086} (\bibinfo{year}{2003}).

\bibitem{davidovitch_spreading_2005}
\bibinfo{author}{Davidovitch, B.}, \bibinfo{author}{Moro, E.} \&
  \bibinfo{author}{Stone, H.~A.}
\newblock \bibinfo{title}{Spreading of viscous fluid drops on a solid substrate
  assisted by thermal fluctuations}.
\newblock \emph{\bibinfo{journal}{Phys. Rev. Lett.}}
  \textbf{\bibinfo{volume}{95}}, \bibinfo{pages}{244505}
  (\bibinfo{year}{2005}).

\bibitem{moseler_nanojets_2000}
\bibinfo{author}{Moseler, M.} \& \bibinfo{author}{Landman, U.}
\newblock \bibinfo{title}{Formation, stability, and breakup of nanojets}.
\newblock \emph{\bibinfo{journal}{Science}} \textbf{\bibinfo{volume}{289}},
  \bibinfo{pages}{1165} (\bibinfo{year}{2000}).

\bibitem{gruen_2006}
\bibinfo{author}{Gruen, G.}, \bibinfo{author}{Mecke, K.} \&
  \bibinfo{author}{Rauscher, M.}
\newblock \bibinfo{title}{Thin-film flow influenced by thermal noise}.
\newblock \emph{\bibinfo{journal}{J. Stat. Phys.}}
  \textbf{\bibinfo{volume}{122}}, \bibinfo{pages}{1261} (\bibinfo{year}{2006}).

\bibitem{eggers_jet2002}
\bibinfo{author}{Eggers, J.}
\newblock \bibinfo{title}{Dynamics of liquid nanojets}.
\newblock \emph{\bibinfo{journal}{Phys. Rev. Lett.}}
  \textbf{\bibinfo{volume}{89}}, \bibinfo{pages}{084502}
  (\bibinfo{year}{2002}).

\bibitem{fetzer_dewetting_2007}
\bibinfo{author}{Fetzer, R.}, \bibinfo{author}{Rauscher, M.},
  \bibinfo{author}{Seemann, R.}, \bibinfo{author}{Jacobs, K.} \&
  \bibinfo{author}{Mecke, K.}
\newblock \bibinfo{title}{Thermal noise influences fluid flow in thin films
  during spinodal dewetting}.
\newblock \emph{\bibinfo{journal}{Phys. Rev. Lett.}}
  \textbf{\bibinfo{volume}{99}}, \bibinfo{pages}{114503}
  (\bibinfo{year}{2007}).

\bibitem{willis_freund_jpcm2009}
\bibinfo{author}{Willis, A.~M.} \& \bibinfo{author}{Freund, J.~B.}
\newblock \bibinfo{title}{Enhanced droplet spreading due to thermal
  fluctuations}.
\newblock \emph{\bibinfo{journal}{J. Phys.: Cond. Mat.}}
  \textbf{\bibinfo{volume}{21}}, \bibinfo{pages}{464128}
  (\bibinfo{year}{2009}).

\bibitem{milchev_binder_jcp2002}
\bibinfo{author}{Milchev, A.} \& \bibinfo{author}{Binder, K.}
\newblock \bibinfo{title}{Droplet spreading: A {M}onte {C}arlo test of
  {T}annerâ€™s law}.
\newblock \emph{\bibinfo{journal}{J. Chem. Phys.}}
  \textbf{\bibinfo{volume}{116}}, \bibinfo{pages}{7691} (\bibinfo{year}{2002}).

\bibitem{gross_fdbe_2011}
\bibinfo{author}{Gross, M.}, \bibinfo{author}{Cates, M.~E.},
  \bibinfo{author}{Varnik, F.} \& \bibinfo{author}{Adhikari, R.}
\newblock \bibinfo{title}{Langevin theory of fluctuations in the discrete
  {B}oltzmann equation}.
\newblock \emph{\bibinfo{journal}{J. Stat. Mech.}}
  \textbf{\bibinfo{volume}{2011}}, \bibinfo{pages}{P03030}
  (\bibinfo{year}{2011}).

\bibitem{gross_dsfd2010}
\bibinfo{author}{Gross, M.}, \bibinfo{author}{Adhikari, R.},
  \bibinfo{author}{Cates, M.~E.} \& \bibinfo{author}{Varnik, F.}
\newblock \bibinfo{title}{Modelling thermal fluctuations in non-ideal fluids
  with the lattice {B}oltzmann method}.
\newblock \emph{\bibinfo{journal}{Phil. Trans. R. Soc. A}}
  \textbf{\bibinfo{volume}{369}}, \bibinfo{pages}{2274} (\bibinfo{year}{2011}).

\bibitem{gross_flbe_2010}
\bibinfo{author}{Gross, M.}, \bibinfo{author}{Adhikari, R.},
  \bibinfo{author}{Cates, M.~E.} \& \bibinfo{author}{Varnik, F.}
\newblock \bibinfo{title}{Thermal fluctuations in the lattice {B}oltzmann
  method for nonideal fluids}.
\newblock \emph{\bibinfo{journal}{Phys. Rev. E}} \textbf{\bibinfo{volume}{82}},
  \bibinfo{pages}{056714} (\bibinfo{year}{2010}).

\bibitem{lee_fischer_2006}
\bibinfo{author}{Lee, T.} \& \bibinfo{author}{Fischer, P.~F.}
\newblock \bibinfo{title}{Eliminating parasitic currents in the {L}attice
  {B}oltzmann equation method for nonideal gases}.
\newblock \emph{\bibinfo{journal}{Phys. Rev. E}} \textbf{\bibinfo{volume}{74}},
  \bibinfo{pages}{046709} (\bibinfo{year}{2006}).

\bibitem{shan_Lattice_1993}
\bibinfo{author}{Shan, X.} \& \bibinfo{author}{Chen, H.}
\newblock \bibinfo{title}{{L}attice {B}oltzmann model for simulating flows with
  multiple phases and components}.
\newblock \emph{\bibinfo{journal}{Phys. Rev. E}} \textbf{\bibinfo{volume}{47}},
  \bibinfo{pages}{1815} (\bibinfo{year}{1993}).

\bibitem{shan_chen_1994}
\bibinfo{author}{Shan, X.} \& \bibinfo{author}{Chen, H.}
\newblock \bibinfo{title}{Simulation of nonideal gases and liquid-gas phase
  transitions by the lattice {B}oltzmann equation}.
\newblock \emph{\bibinfo{journal}{Phys. Rev. E}} \textbf{\bibinfo{volume}{49}},
  \bibinfo{pages}{2941} (\bibinfo{year}{1994}).

\bibitem{swift_prl1995}
\bibinfo{author}{Swift, M.~R.}, \bibinfo{author}{Osborn, W.~R.} \&
  \bibinfo{author}{Yeomans, J.~M.}
\newblock \bibinfo{title}{{L}attice {B}oltzmann simulation of nonideal fluids}.
\newblock \emph{\bibinfo{journal}{Phys. Rev. Lett.}}
  \textbf{\bibinfo{volume}{75}}, \bibinfo{pages}{830} (\bibinfo{year}{1995}).

\bibitem{swift_pre1996}
\bibinfo{author}{Swift, M.~R.}, \bibinfo{author}{Orlandini, E.},
  \bibinfo{author}{Osborn, W.~R.} \& \bibinfo{author}{Yeomans, J.~M.}
\newblock \bibinfo{title}{{L}attice {B}oltzmann simulations of liquid-gas and
  binary fluid systems}.
\newblock \emph{\bibinfo{journal}{Phys. Rev. E}} \textbf{\bibinfo{volume}{54}},
  \bibinfo{pages}{5041} (\bibinfo{year}{1996}).

\end{thebibliography}

\end{document}